\documentclass[]{pasj01} 
\Received{}
\Accepted{}
\usepackage[switch,mathlines]{lineno}
\usepackage{graphicx}	
\usepackage{bm}  
\usepackage{color} 
\usepackage[T1]{fontenc} 
\DeclareRobustCommand{\VAN}[3]{#2}
\let\VANthebibliography\thebibliography
\def\thebibliography{\DeclareRobustCommand{\VAN}[3]{##3}\VANthebibliography}  

\def\be{\begin{equation}}\def\ee{\end{equation}}
\def\vlsr{v_{\rm LSR}} \def\Msun{M_\odot} \def\deg{^\circ}  
\def\co{$^{12}$CO } \def\Xco{X_{\rm CO}}   \def\Tb{T_{\rm B}} 
\def\Htwo{H$_2$ }   \def\Kkms{K km s$^{-1}}   \def\kms{km s$^{-1}$}   \def\Kkms{K \kms } 
   \def\Tb{T_{\rm B}}        \def\ekms{{\rm km\ s^{-1}}}  \def\epc{{\rm pc} } 
 \def\apj{ApJ} \def\aap{AA} \def\mnras{MNRAS} \def\pasj{PASJ} \def\aj{AJ}
 
 \def\log{{\rm log}} 
 \def\tc{t_{\rm C}}   
\def\tc{t_{\rm c}}\def\lc{l_{\rm c}}\def\vc{v_{\rm c}}\def\rc{r_{\rm c}}\def\nc{n_{\rm c}}  
\def\Msun{M_{\odot \hskip-5.2pt \bullet}}    \def\kms{km s$^{-1}$}  \def\deg{^\circ}   \def\Htwo{H$_2$\ }     
 \def\sfuvol{\Msun~{\rm y^{-1}~kpc^{-3}}}
\def\log{{\rm log}}

\def\bc{\begin{center}}\def\ec{\end{center}}

\def\rc{r_{\rm c}} \def\vc{v_{\rm c}}

\def\nc{n_{\rm c}}
\def\redddd{}
\def\reddd{}
\def\redd{}
\def\red{}

\def\ekms{{\rm km\ s^{-1}}}
\def\rhosfr{\rho_{\rm SFR}} 
 
\def\rhohtwo{\rho_{\rm H_2}}
\def\HII{H{\sc II} }  

\begin{document}   

\title{Stochastic star formation in the Milky Way inferred from the unity index of Kennicutt-Schmidt law}
\author{Yoshiaki \textsc{Sofue}\altaffilmark{1} }
\altaffiltext{1}{Institute of Astronomy, The University of Tokyo, Mitaka, Tokyo 186-0015, Japan}
\email{sofue@ioa.s.u-tokyo.ac.jp}

 
\maketitle 

\begin{abstract}    
We performed a correlation analysis between the brightness temperature of the CO line and number density of \HII regions in the longitude--velocity diagram (LVD) of the Milky Way in order to investigate the volumetric star-formation law.
We determined the index $\alpha$ of the Kennicutt-Schmidt (KS) law for the molecular gas defined by $\rhosfr \propto \rho_{\rm H_2}^\alpha$, where $\rhosfr$ is the SFR (star-formation rate) density and $\rhohtwo$ is the molecular-gas density. 
We obtained $\alpha= 1.053 \pm 0.075$ and $1.031 \pm 0.067$ for the CO-line data from the Nobeyama 45-m and Columbia 1.2-m telescope Galactic plane surveys, respectively. 
This result is consistent with the KS indices currently determined for the molecular gas in the Milky Way as well as in spiral and starburst galaxies.
We argue that an index close to 1 is universal in favour of stochastic (spontaneous) star formation, but is inconsistent with cloud-collision model, which predicts a steeper index of $\alpha=2$. 
We also suggest that the efficiency of star formation in the Galactic Centre is an order of magnitude lower than that in the disc.
\end{abstract}    

\section{Introduction}  

The trigger for star formation (SF) in molecular clouds in the Galaxy has been one of the fundamental subjects of the ISM over recent decades \citep{1987ARA&A..25...23S,2003ARA&A..41...57L}.
There are two major ideas: 
one scenario is the stochastic (spontaneous) SF by self-regulation mechanisms in individual molecular clouds due to the gravitational instability and/or sequential compression by expanding shells of \HII regions \citep{1977ApJ...214..725E,2009ApJ...700.1609M,2023ASPC..534..153H,2023ASPC..534..233P}.
The other scenario is the external triggering by collisions of molecular clouds
\citep{1992PASJ...44..203H,1994ApJ...429L..77H,1996A&A...308..979K,2021PASJ...73S...1F}. 
 
A possible way to clarify the SF mechanisms is to check the index $\alpha$ of the Kennicutt-Schmidt (KS) law \citep{2012ARA&A..50..531K}
\reddd{defined through  
\begin{equation}   \
\rhosfr =A\ \left({\rhohtwo}\right)^{\alpha},
 \end{equation}    
where $\rhosfr \ [\sfuvol]$ and $\rhohtwo $ [H$_2$ cm$^{-3}$] are the volumetric densities of SFR and the molecular gas, respectively, and $A$ is a constant.}
Stochastic SF predicts $\alpha = 1$, because $\rhosfr$ is proportional to the density of the clouds, whereas the collision process requires $\alpha =2$, because $\rhosfr$ depends on the collision frequency between clouds.
The current measurements have shown $\alpha\sim 1$ in the Milky Way \citep{2009AJ....137..266F,2017MNRAS.469.1647S,2017PASJ...69...19S,2022ApJ...941..162E} 
and spiral galaxies \citep{2006PASJ...58..793K,2012ARA&A..50..531K},
indicating SFR density more linearly proportional to the molecular-gas density.

\red{The KS law in the Milky Way has been studied using the face-on maps of the molecular and HI gases and of \HII regions \citep{2017PASJ...69...19S,2017MNRAS.469.1647S,2019A&A...622A..64B,2021A&A...653A..63S,2022ApJ...941..162E}.}
Recent studies of the face-on transformation (FOT) from the radial velocity to line-of-sight distance have shown that the derived gas distribution is highly sensitive to the rotation curve (RC) \citep{2017A&A...607A.106M,2023PASJ...75..279F}.
So, in this paper, we propose a new, simpler and more direct method to derive the KS-law index using the longitude--velocity diagram (LVD) without employing the FOT \citep{2016ApJ...823...76K}. 
By applying the new method, we determine the KS law index $\alpha$ for the molecular gas, and discuss the feasibility of the stochastic and collision models for the star formation. 
 
In the analysis, we use the archival data of \co-line emission from the FUGIN (the Four-receiver system Unbiased Galactic plane Imaging survey with the Nobeyama 45-m telescope) \citep{2017PASJ...69...78U}, Galactic Centre survey \citep{2019PASJ...71S..19T}, and Columbia 1.2-m Galactic plane survey \citep{2001ApJ...547..792D},
combined with the WISE (Wide-field Infrared Survey Explore) \HII region catalogue \citep{2014ApJS..212....1A}.

\section{Molecular-gas KS law using LVD} 

\begin{figure}   
\begin{center}   
\includegraphics[width=8cm,height=7.5cm]{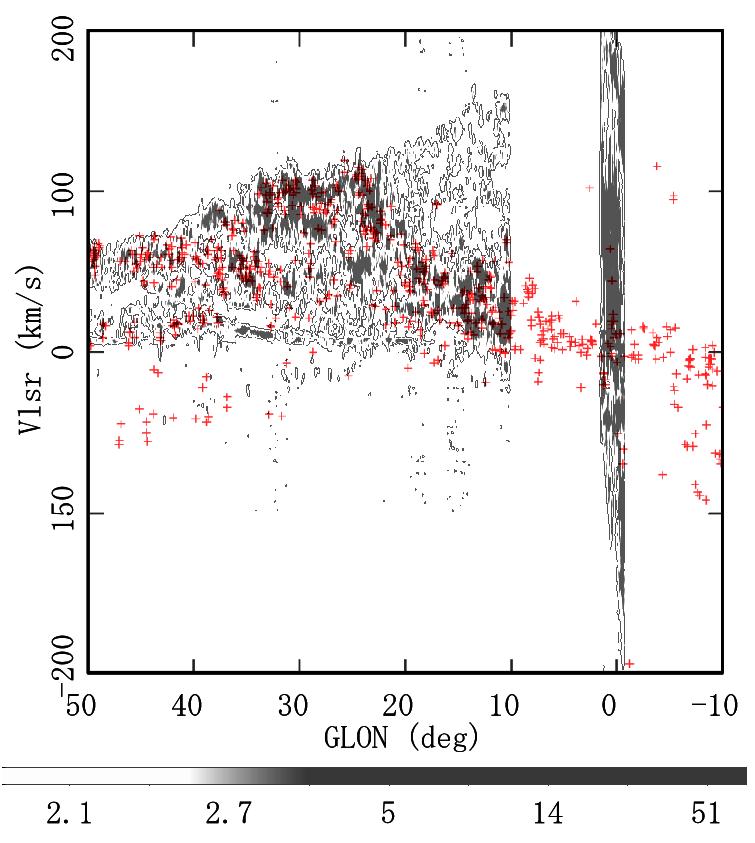}  
\end{center}
\vskip -2mm
\caption{ 
HII regions in the Galactic plane at $|b|\le 0\deg.2$ from the WISE catalogue  \citep{2014ApJS..212....1A} superposed on the LVD of CO-line brightness constructed from the data by the Nobeyama 45-m FUGIN \citep{2017PASJ...69...78U} and GC survey \citep{2019PASJ...71S..19T}.
The bar indicates $\Tb$ inK and the contours are drawn at $\Tb=0.5$ K.
}
\label{fig1}	
\end{figure}

\redd{We perform correlation analysis on the LVD between the number density of \HII regions from the WISE catalogue \citep{2014ApJS..212....1A} 
and the volume density of H$_2$ molecules calculated from the Nobeyama CO-line surveys \citep{2017PASJ...69...78U,2019PASJ...71S..19T} in the inner disc of the Milky Way.
The correlation on the LVD is more directly coupled to the data, bypassing the sophisticated face-on transition procedures \cite{2016ApJ...823...76K}.
Figure \ref{fig1} plots the longitude--velocity positions of \HII regions in  the Galactic plane superposed on the CO-line LVDs along the Galactic plane ($b=0\deg$).}
As readily known \citep{2014A&A...569A.125H}, \HII regions and CO-line intensity are tightly correlated in the LVD.

\def\rtan{R_{\rm tan}}

{In our study we constrain the region of analysis in the Galactic plane by using only the \HII regions in the Galactic plane at $|b|\le 0\deg.2$ and CO LVD at $b=0\deg$.
The region was so chosen in order to avoid the uncertainty during the determination process of the height from the Galactic plane depending on the distance ambiguity.
Thus we selected 492 \HII regions, about 38\% of the 1316 catalogued sources with measurements of LSR velocities in the radio recombination lines (RRL).}
We then calculate the number density $N$ of \HII regions and CO brightness temperature $\Tb$ averaged in every $(l,v)$ grid with bin size $\delta l \times \delta v = 0\deg.25 \times 10$ \kms for Nobeyama and $2\deg\times 5$ \kms for Columbia CO survey data.

Using the thus-obtained sets of $N$ and $\Tb$, we calculate the volume densities $n_i$ of \HII regions ($i=$ HII) and H$_2$ molecules ($i=$ H$_2$) by
\be
n_i=\frac{dN_i}{ds}=\frac{dN_i}{dv}\frac{dv}{ds},
\ee
where  $v=\vlsr$ is the LSR radial velocity and $s$ is the distance from the Sun.
Recalling 
$N_{\rm H_2}=\Xco \int \Tb dv$, 
we have
\be
n_{\rm H_2}=\Xco \Tb (dv/ds),
\ee
where $\Xco$ is the CO-to-\Htwo conversion factor assumed to be constant at $2\times 10^{20}$ \Htwo cm$^{-3}$ [\Kkms]$^{-1}$ \citep{koh23}.  
The velocity gradient, $dv/ds$, is calculated by 
\be
dv/ds=(dV/dR-V/R)k\sqrt{1-k^2},
\ee
where $R$ is the galactocentric radius, $V=V(R)$ is the rotation velocity, $k=R_0 \sin l/R$, and $v$ and $R$ are related by
$  v=(V {R_0}/{R} - V_0 ) \sin l$.  
We adopt the most accurate rotation curve of the Milky Way \citep{2021PASJ...73L..19S,2023Ap&SS.368...74S}. 

Then, we calculate the following values using the LVD:
\be
n^*_{\rm \HII}=(dN_{\rm \HII}/dv)(dv/ds)
\label{eq_nhii}
\ee
and
\be
n^*_{\rm H_2}=n_{\rm H_2}/\Xco=\Tb (dv/ds).
\label{eq_nhtwo}
\ee
\reddd{These quantities are assumed to be proportional to $\rhosfr$  and  $\rhohtwo$, respectively, averaged in each bin with size $\delta \vlsr \times \delta l$. In order to check the index $\alpha$, we hereafter investigate the correlation between log $n^*_{\rm \HII}$ and log $n^*_{\rm H_2}$.}
Except for the scaling, the plots are equivalent to the KS plot.  

\begin{figure*}  
\begin{center}   
A\includegraphics[width=6cm]{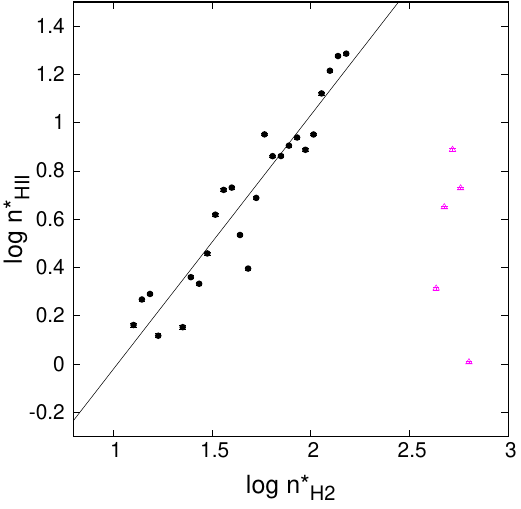} 
B\includegraphics[width=6cm]{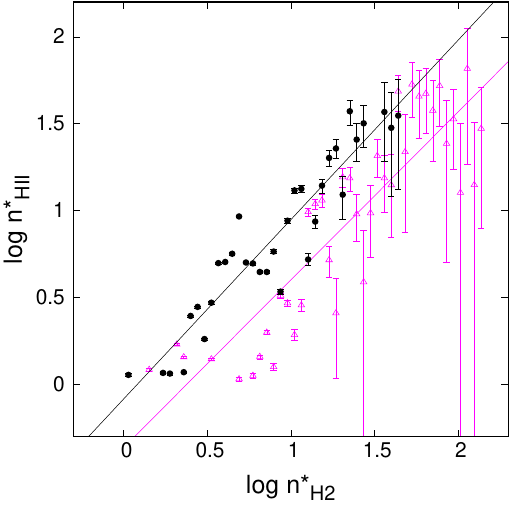}\\
\vskip 2mm
C\includegraphics[width=6cm]{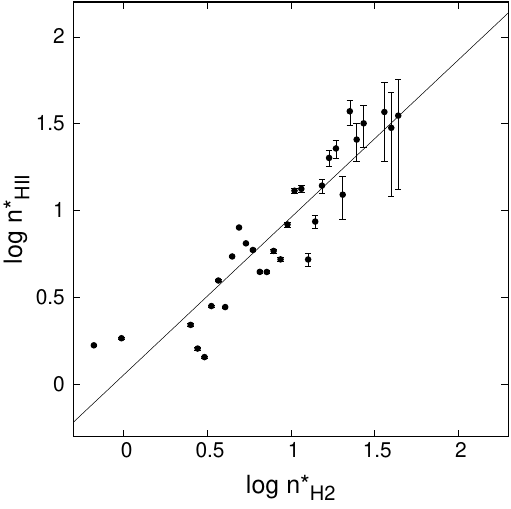}
D\includegraphics[width=6cm]{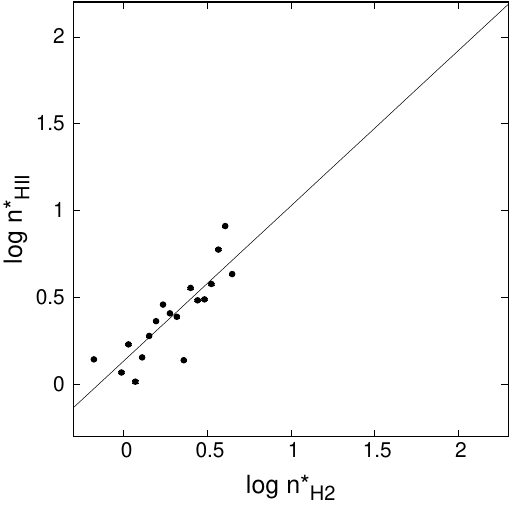}  
\end{center}
\vskip -2mm
\caption{Plots of log $n_{\rm HII}^*$  using the WISE \HII region catalogue against log $n_{\rm H_2}^*$ using
[A] the Nobeyama 45-m \co\ data at $|l|=10\deg-50\deg$ (black dots) and GC $|l|\le 5\deg$ (magenta). 
The grid size is  $\delta l \times \delta v=0.2\deg \times 10$ \kms.
[B] Same, but using the Columbia 1.2 m \co\ at $|l|=10\deg-180\deg$ (black) and $|l| \le 10\deg$ (magenta). 
The grid size is $2\deg\times 5 \ekms$. 
[C] Columbia $|l|=10\deg-30\deg$ inside 4 kpc. 
[D] Columbia $|l| =30\deg-180\deg$ outside 4 kpc.  
Bars are standard errors in the original scaling data in the 0.05 dex bins of $n^*_{\rm H_2}$.    
\redd{The straight lines are linear least-squares fit to the log-scaling plots with equal weighting.}
}
\label{fig2}	
\end{figure*}

Figure \ref{fig2} shows the thus-obtained plots of log $n^*_{\rm \HII}$ against log $n^*_{\rm H_2}$ averaged in 0.05--0.1 dex bins of the horizontal axes, where the bars are standard errors (SE).
We used the data exceeding thresholds of $n^*_{\rm HII}> n^*_{\rm \HII; min}=1$ and $n^*_{\rm H_2}>n^*_{\rm H_2; min}= 0.1$ in order to avoid the lowest-count data.  
We then determined the KS index $\alpha$ by the least-squares fitting by a linear line in the log--log plane, putting equal weighting to the points, which gave relatively larger weights to higher-density regions where the number of data points gets smaller so that the SE gets larger.
We obtained almost the same results even if we used weights proportional to the inverse of the error bars. 
The fitting results are shown by the thin lines, and we obtain
$\alpha= 1.053 \pm 0.075$ for the Nobeyama data in the inner disc at $50\deg \ge l \ge 10\deg$ (figure \ref{fig2}, panel A), and 
$ 1.031 \pm  0.067$ for Columbia data in the entire disc avoiding the central region at $|l|\ge 10\deg$ (panel B). 
We find nearly equal value 
$\alpha= 0.905 \pm 0.072$ inside the 4 kpc molecular ring at $30\deg \ge |l|\ge 10\deg$ (panel C) for the Columbia data, and a slightly smaller value 
$ 0.892 \pm  0.138$ outside 4 kpc (panel D).

The Nobeyama data in the GC at $|l|\le 2\deg$ (panel A, magenta crosses) show significantly lower number density of \HII regions despite much higher gas density, showing an order of magnitude lower efficiency of star formation.
Accordingly, the Columbia data in the central region at $|l|\le 10\deg$ (magenta crosses and line in panel B), which are mixture of the GC and Galactic disc values, indicate lower \HII density but with almost the same index of $\alpha=0.965 \pm  0.087$.
The larger number of data points in panel B despite the narrower longitude range is due to the steeply increasing velocity range toward the GC.
 
We emphasize that the present method uses LVDs directly, not intervened by the sophisticated FOT procedure to convert the radial velocity to the distance, which includes the difficulty of discriminating far- and near-side solutions.
Therefore, the measured quantities are more reliable compared to those using the current methods insofar as the power-law index of the KS law is concerned.
In table \ref{tab-index} we compare the result with the current determination for the Milky Way and spiral and starburst galaxies.

\redd{The volumetric index obtained here is slightly larger than that from the current FOT method yielding $\alpha =0.7$--0.8 \reddd{(table \ref{tab-index} lines 6 and 7).}
Also, the surface index for the Milky Way lie at slightly larger values of $\sim 1.15$ \reddd{(lines 8--10)}.
Besides the Milky Way, extensive studies about the KS law have been obtained in spiral galaxies also indicating mild indices of $\alpha \simeq 0.8-1.3$, which are consistent with the present new values.   
Therefore, we may state that the unity KS-law index is universal.
For reference we calculated a simple mean of the listed values except for lines 3 and 4 in table \ref{tab-index} as $\left<\alpha \right>=1.048\pm 0.182$.}

On the other hand, a steeper index around $\alpha \sim 1.5$ has been obtained for total gas including HI in galaxies \citep{1998ApJ...498..541K,2021ApJ...908...61K} and in the Milky Way \citep{2017MNRAS.469.1647S,2021PASJ...73L..19S}.
Analysis of the total surface gas density and SFR density from M dwarf stars also yielded $\alpha_{\rm sur} \sim 1.4$ \citep{2009AJ....137..266F}. 
\reddd{Also, an even higher values of $\alpha \gtrsim 2$ are reported for the total gas KS law \citep{2003MNRAS.346.1215B,2006A&A...459..113M}}.
The difference between the indices of the molecular and total (\Htwo+HI) gas KS laws would be due to the intervening phase transition from HI to \Htwo, and vice versa. 
A detailed analysis of the correlation between HI and \Htwo will be the subject for a separate paper.

\begin{table*} 
\caption{Molecular-gas KS index $\alpha$ in the Milky Way \reddd{(first block) compared with those from the literature (following blocks)}. }   
\begin{tabular}{llll}  
\hline 
\hline
Object & Index $\alpha$ & Method$^\dagger$ &Remarks \\  
\hline  
\hskip -3mm {\bf Milky Way (this work)}\\
Inner disc ($|l|=10\deg$--$50\deg$)& $  1.053    \pm   0.075$ &Vol., LVD, Nobe.45m+WISE$^\ddagger$ &Fig.\ref{fig2} panel A\\
Entire disc $(|l|\ge 10\deg)$& $1.031 \pm 0.067$ &Vol., LVD, Col.1.2m+WISE &ibid panel B\\
~~~Inside 4-kpc$^*$ ($|l|=10\deg$--$30\deg)$& $ 0.905 \pm      0.072$ &Vol., LVD, ibid &ibid panel C\\
~~~Outside 4-kpc$^*$ $(|l|> 30\deg)$& $ 0.892   \pm    0.138$ &Vol., LVD, ibid &ibid panel D\\
Inside $|l|<10\deg$ & $  0.965    \pm   0.087$ &Vol., LVD, ibid &ibid panel B\\
\hline 
\hskip -3mm {\bf Milky Way (literature)}\\
Entire disc& $0.78\pm0.05$ &Vol., 3D, Col.1.2m+Hou.$^\dagger$ 
&\citet{2017MNRAS.469.1647S}\\ 
ibid& $0.73$ & Vol. $\rhosfr$& \citet{2019A&A...622A..64B}\\ 
ibid & $1.12\pm0.05$ & Surf. 3D, Col.1.2m+Hou &\citet{2017MNRAS.469.1647S}\\ 
ibid & $1.14\pm 0.07$ & Surf., face on& \citet{2022ApJ...941..162E}\\
Solar vicinity 2 kpc & $1.19\pm 0.09$&Surf., local clouds \& \HII &Fit$^\#$ to fig.18 of \citet{2021A&A...653A..63S}\\  
   \hline 
\hskip -3mm {\bf Galaxies (literature)}\\
NGC/DDO & 1.081& Volume&\citet{1998ApJ...498..541K}\\ 
  NGC/starburst & $1.4\pm 0.15$&Surf.&\citet{2021ApJ...908...61K}\\ 
  NGC/DDO  & 0.925& Surf.&\citet{2023MNRAS.518.4024D}\\%
  NGC high mol. den. & $1.33\pm 0.08$&Surf.&\citet{2006PASJ...58..793K}\\ 
  UGC &  $0.99\pm 0.08$&Surf.&Komugi et al. (2005, 2012)\\
  NGC/IC/Mrk/Arp& $1.0$ & Surf. mol. CO/HCN/IR&\citet{2004ApJ...606..271G}\\
  Disc galaxies & $1.0\pm 0.15$ &Surf, ALMA CO&\citet{2013AJ....146...19L}\\
  Sub-mm galaxies &0.81 - 0.84 &Surf, sub-mm &\citet{2017A&A...602L...9M}\\
  Interacting galaxies&$1.3\pm 0.04$&Surf, CO obs. &\citet{2022PASJ...74..343K}\\ 
  M83 &$1.0\pm 0.30$&Surf. ALMA CO &Fit$^\#$ to fig.14 of \citet{hir18}\\ 
{Mol.-rich galaxies} &  1.07$^{**}$  &Surf, CO, $A_{\rm v}$ cor. H$\alpha$&\citet{2002ApJ...569..157W}\\ 
\hline 
Simple mean of all listed $\alpha$&$1.048\pm 0.182$ &Equal weighting& Except lines 3, 4 for duplication$^*$\\
  \hline    
\end{tabular}  
\begin{tabnote}
$^\dagger$ Columbia 1.2 m \citep{2001ApJ...547..792D},
Hou et al catalogue \citep{2009A&A...499..473H},
$^\ddagger$ Nobeyama 45 m FUGIN and GC surveys \citep{2017PASJ...69...78U,2019PASJ...71S..19T},
WISE\citep{2014ApJS..212....1A}:
$^\#$ $\alpha$ of the least-squares fit to the plots by $\log\ \Sigma_{\rm SFR}=\alpha \log\ \Sigma_{\rm gas}$ with equal weighting.
$^*$ Lines 3 and 4 are not used for the mean in the last row.
$^{**}$ Average of $0.78\pm  0.34$ with constant $A_{\rm v}$ correction for H$\alpha$ and $1.36\pm  0.35$ with $A_{\rm v}$ proportional to H density. 
\end{tabnote}
\label{tab-index}
\end{table*}

\section{Discussion}

\subsection{Comparison with KS laws by other methods}

The 3D molecular gas maps of the Milky Way derived in our earlier works 
based on the Columbia CO survey and Hou's \HII region catalogue \citep{2009A&A...499..473H} made it possible to measure both the volume and surface densities of the \HII regions and molecular gas \citep{2017MNRAS.469.1647S,2017PASJ...69...19S}.  
\reddd{In these works, we used FOT maps and obtained a smaller slope of $\alpha=0.78\pm 0.05$ with $\log\ A=-2.6\pm 0.05$ for the volume density, while a larger value $\alpha=1.12\pm 0.05$ for surface densities.
However, a more recent work on the surface-density led to a value closer to unity \citep{2022ApJ...941..162E}.}
In the meantime, during the derivation of face-on molecular-gas maps of the inner Milky Way
\citep{2023MNRAS.525.4540S} based on the most accurate rotation curve \citep{2021PASJ...73L..19S},
it was shown that the kinematic distance is highly sensitive to the adopted rotation curve.
This often results in erroneous maps associated with artifact holes and/or over-condensation of gas near and along the tangent-point circle.
It was also shown that the maps are strongly disturbed by the non-circular motion such as due to the 3 kpc expanding ring.
Because these problems significantly disturb the KS-law analyses using the face-on maps,
we here proposed a more direct method to derive the KS law using the LVDs without employing the FOT.

\subsection{Cloud-collision model with $\alpha=2$}

We here consider the cloud-collision process and expected KS law.
Let the radius of colliding clouds be $\rc$, number density $\nc$, and velocity dispersion of the clouds $\vc$.
Then, we have the mean free path 
 \be  
\lc = (\nc \pi \rc ^2)^{-1} 
\ee  
and collision time
\be
\tc= \lc/\vc =(\vc \nc \pi \rc ^2)^{-1}.  
\ee 
The corresponding SFR density is calculated by 
\be
\rhosfr=\eta M_{\rm c}\nc \tc^{-1} = \pi \eta M_{\rm c} \nc^2 \rc ^2 \vc \propto \nc^2,
\ee 
where $\eta$ is the SF efficiency per one cloud collision and $M_{\rm c}$ is typical mass of molecular clouds
\footnote{The same formulation applies to surface density $\Sigma$ by replacing $\pi \rc^2$ with $2\rc$, $\nc$ with $\Sigma_{\rm c}$, yielding $\Sigma_{\rm SFR}\propto \Sigma_{\rm  c}^2$}.

\reddd{Thus, we obtain $\alpha=2$, $\lc\sim 50$ kpc, $\tc\sim 3$ Gyr, and $ A\sim 10^{-3}\eta $,
if we adopt the largest size ($2\rc=5$ pc) of the estimated typical cloud sizes (1--5 pc) and collision velocity $\vc\sim 15$ \kms of the estimated 10--15 \kms by the collision hypothesis \citep{2021PASJ...73S...1F}.
Here, we assumed \redddd{$M_{\rm c}\sim 10^{4-5} \Msun$} and 
$\nc\sim (100 ~\epc)^{-3} $ referring to the molecular cloud density in the Galactic disc \citep{1987ApJ...319..730S}.
Thus obtained slope does not fit the observation, and the amplitude is too small compared with that in the Milky Way ($A\sim 10^{-3}$), unless an anomalously high SF efficiency of $\eta\sim 1$ is attained.} 

On the other hand, a number of pieces of 'evidence' for cloud collisions are reported \citep{2021PASJ...73S...1F} (and the literature therein).
As to the morphology of cavities in the clouds and position-velocity ridges, we point out that both can be created by an expanding molecular shell driven by an expanding \HII region \citep{2023MNRAS.525.4540S}. 
We also mention that two unbound objects moving on hyperbolic orbits rarely collide, unless the angular momentum is removed. 
So far, there seems to be no measurement of the angular momentum or transverse velocity, line-of-sight distance, and the sense of approaching or receding. 
Also, the reported mutual velocities, $\sim 10$--30 \kms, allow for kinematic distance of $\sim 1$ kpc between the clouds according to the Galactic rotation.   
So, parts of the evidence do not seem to be conclusive yet, leaving room to consider different SF triggering mechanisms.

\subsection{Stochastic star formation with $\alpha=1$}

We have shown that the KS-law index of the Milky Way is unity, $\alpha \simeq 1.0$, which indicates the stochastic SF scenario.   
A possible mechanism for stochastic SF is the self-gravitational contraction of isolated molecular clouds, which works in the scheme of the sequential star formation \citep{1977ApJ...214..725E} 
and gravitational instability in the hub and filaments 
\citep{2009ApJ...700.1609M}.
The growth (Jeans) time of instability in a molecular cloud of density $\sim 10^3 - 10^4$ H$_2$ cm$^{-3}$ is $t_{\rm J}\sim 1/\sqrt{4 \pi G\rho}\sim 0.2-0.5$ Myr. 
If this time is regarded as the SF time, the stochastic SF is more efficient than that presumed by cloud collision time as estimated in the previous subsection. 

The global SFR density $\rhosfr$ is proportional to the number density $\nc$  of clouds in the averaging bins with $\delta l\times \delta v$, where the shape of the mass function of molecular clouds is assumed to be universal and the Jeans time of individual clouds with the same mass is equal to each other. 
On the other hand, the probability of SF in each cloud by gravitational collapse depends on the gas density and velocity dispersion inside each cloud, but not on  the environment. The SFR density averaged in an area sufficiently wider than clouds' size is, therefore, simply proportional to the number density of clouds, or $\alpha=1$, which is indeed observed in the present analysis.

\subsection{Suppression of SF in the Galactic Centre}

Figure \ref{fig2} showed that the SFR density in the GC is comparable to that in the disc despite the much higher density of molecular gas.
This indicates a significantly lower efficiency of star formation, in agreement with the current estimation of an order of magnitude lower star-formation efficiency \citep{2014MNRAS.440.3370K,2017MNRAS.469.2263B}.  
The SF in the GC is, thus, possibly suppressed by some external actions such as high magnetic pressure, fast differential rotation with strong shear, turbulence induced by explosive events in the nucleus, and/or external disturbance by falling gaseous debris from the merged companions. 
We point out that the low SF efficiency (SFE) in the GC despite of the high density and high velocity dispersion does not support the collision scenario. 

\reddd{However, we comment that one cannot ignore the possibility that the apparently low SFE is merely due to the detection limit of the current RRL observations beyond $\sim 8$ kpc particularly in the GC direction for the high background emission.}
  
\subsection{Limitation of the analysis}

We used the CO-line LVD along the Galactic plane at $b=0\deg$ for the molecular gas in order to represent the densest part of the disc and to avoid the complexity arising from the mixture of gases at different distances and heights. 
Accordingly, we constrained the \HII regions to near Galactic plane objects at $|b|\lesssim 0\deg.2$, typically $h\lesssim 20$ pc, by which about half of the catalogued \HII regions with RRL velocities were not used. 
This gave fewer \HII regions compared to the current studies, while it avoided the uncertainty arising from uncertain distances and heights.

\redd{Another concern is the incompleteness of the catalogued \HII regions caused by the detection limit of the RRL due to decreasing fluxes with the distance.
This yields apparently decreasing density of \HII regions beyond several kpc \citep{2014A&A...569A.125H}.  
On the other hand the detection limit of the CO line does not depend on the distance insofar as the molecular disc is resolved. 
This situation results in missing \HII regions for finite CO intensity regions beyond several kpc, so that the SFR density there is underestimated, and may cause a systematic error in the slope.} 

\redd{We checked this point by dividing the data into different longitudinal ranges, so that they represent data sets with different mean distances from the observer according to the lopsided location of the Sun in the Galaxy.
As shown in table \ref{tab-index}, we do not find significant dependence of the index on the regions.
So, we consider here that the effect is rather even or negligible, not strongly disturbing the general property of KS index.
However, in order to ultimately answer this question \reddd{and the reason for the low SFE in the GC}, we may need to wait for an RRL survey sufficiently sensitive to cover the entire Galaxy. }

\section{Summary}

By correlation analysis of the distributions of molecular gas and \HII regions in the longitude--velocity diagrams of the Milky Way, we determined the KS index to be  $\alpha\simeq 1.0$  in the Galaxy.
The index is consistent with those derived in spiral galaxies for molecular gas as listed in table \ref{tab-index}.
The unity KS index favours stochastic (spontaneous) star formation by self-regulation in individual molecular clouds, while it contradicts the cloud-collision model. 
We also suggested that the SFR density in the Galactic Centre may be lower than that in the disc by an order of magnitude, indicating anomalous suppression of SF.

\section*{Acknowledgments} 
The data analysis was performed at the Astronomical Data Analysis Center of the National Astronomical Observatories of Japan.
The author thanks Drs. T. Umemoto (FUGIN), T. Oka (Nobeyama GC survey), T. Dame (Columbia CO survey), and L. Anderson (WISE) for making the archival data available for us.

\section*{Data availability}
The Colombia CO data were retrieved from the URL:
{https:// lweb.cfa.harvard.edu/ rtdc/ CO/};
The WISE \HII region catalogue: 
{http:// astro.phys.wvu.edu/wise/};
FUGIN Nobeyama CO survey:  
{http://jvo.nao.ac.jp/portal};
and GC CO survey: 
{https:// www.nro.nao.ac.jp/ $\sim$nro45mrt/html/ results/data.html} 

\section*{Conflict of interest}
There is no conflict of interest.



\begin{thebibliography}{}   
\bibitem[Anderson et al.(2014)]{2014ApJS..212....1A} Anderson, L.~D., Bania, T.~M., Balser, D.~S., et al.\ 2014, \apjs, 212, 1. 
\bibitem[Bacchini et al.(2019)]{2019A&A...622A..64B} Bacchini, C., Fraternali, F., Iorio, G., et al.\ 2019, \aap, 622, A64. 
\bibitem[Barnes et al.(2017)]{2017MNRAS.469.2263B} Barnes, A.~T., Longmore, S.~N., Battersby, C., et al.\ 2017, \mnras, 469, 2263. 
\bibitem[Boissier et al.(2003)]{2003MNRAS.346.1215B} Boissier, S., Prantzos, N., Boselli, A., et al.\ 2003, \mnras, 346, 1215. 
\bibitem[Dame et al.(2001)]{2001ApJ...547..792D} Dame, T.~M., Hartmann, D., \& Thaddeus, P.\ 2001, \apj, 547, 792.
\bibitem[Du et al.(2023)]{2023MNRAS.518.4024D} Du, K., Shi, Y., Zhang, Z.-Y., et al.\ 2023, \mnras, 518, 4024. 
\bibitem[Elia et al.(2022)]{2022ApJ...941..162E} Elia, D., Molinari, S., Schisano, E., et al.\ 2022, \apj, 941, 162. 
\bibitem[Elmegreen \& Lada(1977)]{1977ApJ...214..725E} Elmegreen, B.~G. \& Lada, C.~J.\ 1977, \apj, 214, 725. 
\bibitem[Fuchs et al.(2009)]{2009AJ....137..266F} Fuchs, B., Jahrei{\ss}, H., \& Flynn, C.\ 2009, \aj, 137, 266. 
\bibitem[Fujita et al.(2023)]{2023PASJ...75..279F} Fujita, S., Ito, A.~M., Miyamoto, Y., et al.\ 2023, \pasj, 75, 279. 
\bibitem[Fukui et al.(2021)]{2021PASJ...73S...1F} Fukui, Y., Habe, A., Inoue, T., et al.\ 2021, \pasj, 73, S1.
\bibitem[Gao \& Solomon(2004)]{2004ApJ...606..271G} Gao, Y. \& Solomon, P.~M.\ 2004, \apj, 606, 271. 
\bibitem[Habe \& Ohta(1992)]{1992PASJ...44..203H} Habe, A. \& Ohta, K.\ 1992, \pasj, 44, 203
\bibitem[Hacar et al.(2023)]{2023ASPC..534..153H} Hacar, A., Clark, S.~E., Heitsch, F., et al.\ 2023, 
534, 153. 
\bibitem[Hasegawa et al.(1994)]{1994ApJ...429L..77H} Hasegawa, T., Sato, F., Whiteoak, J.~B., et al.\ 1994, \apjl, 429, L77. 
\bibitem[Hirota et al.(2018)]{hir18}
Hirota, A., Egusa, F., Baba, J., et al.\ 2018, \pasj, 70, 73. 
\bibitem[Hou et al.(2009)]{2009A&A...499..473H} Hou, L.~G., Han, J.~L., \& Shi, W.~B.\ 2009, \aap, 499, 473. 
\bibitem[Hou \& Han(2014)]{2014A&A...569A.125H} Hou, L.~G. \& Han, J.~L.\ 2014, \aap, 569, A125. 
\bibitem[Kaneko et al.(2022)]{2022PASJ...74..343K} Kaneko, H., Kuno, N., Iono, D., et al.\ 2022, \pasj, 74, 343. 
\bibitem[Kennicutt(1998)]{1998ApJ...498..541K} Kennicutt, R.~C.\ 1998, \apj, 498, 541. 
\bibitem[Kennicutt \& De Los Reyes(2021)]{2021ApJ...908...61K} Kennicutt, R.~C. \& De Los Reyes, M.~A.~C.\ 2021, \apj, 908, 61. 
\bibitem[Kennicutt \& Evans(2012)]{2012ARA&A..50..531K} Kennicutt, R.~C. \& Evans, N.~J.\ 2012, \araa, 50, 531. 
\bibitem[Kimura \& Tosa(1996)]{1996A&A...308..979K} Kimura, T. \& Tosa, M.\ 1996, \aap, 308, 979
\bibitem[Koda et al.(2016)]{2016ApJ...823...76K} Koda, J., Scoville, N., \& Heyer, M.\ 2016, \apj, 823, 76. 
\bibitem[Kohno \& Sofue(2023)]{koh23}
Kohno, M. \& Sofue, Y.\ 2023, arXiv:2311.13760. 
\bibitem[Komugi et al.(2006)]{2006PASJ...58..793K} Komugi, S., Sofue, Y., \& Egusa, F.\ 2006, \pasj, 58, 793. 
\bibitem[Komugi et al.(2005)]{2005PASJ...57..733K} Komugi, S., Sofue, Y., Nakanishi, H., et al.\ 2005, \pasj, 57, 733. 
\bibitem[Komugi et al.(2012)]{2012ApJ...757..138K} Komugi, S., Tateuchi, K., Motohara, K., et al.\ 2012, \apj, 757, 138. 
\bibitem[Kruijssen et al.(2014)]{2014MNRAS.440.3370K} Kruijssen, J.~M.~D., Longmore, S.~N., Elmegreen, B.~G., et al.\ 2014, \mnras, 440, 3370. 
\bibitem[Lada \& Lada(2003)]{2003ARA&A..41...57L} Lada, C.~J. \& Lada, E.~A.\ 2003, \araa, 41, 57. 
\bibitem[Leroy et al.(2013)]{2013AJ....146...19L} Leroy, A.~K., Walter, F., Sandstrom, K., et al.\ 2013, \aj, 146, 19. 
\bibitem[Marasco et al.(2017)]{2017A&A...607A.106M} Marasco, A., Fraternali, F., van der Hulst, J.~M., et al.\ 2017, \aap, 607, A106. 
\bibitem[Miettinen et al.(2017)]{2017A&A...602L...9M} Miettinen, O., Delvecchio, I., Smol{\v{c}}i{\'c}, V., et al.\ 2017, \aap, 602, L9. 
\bibitem[Misiriotis et al.(2006)]{2006A&A...459..113M} Misiriotis, A., Xilouris, E.~M., Papamastorakis, J., et al.\ 2006, \aap, 459, 113. 
\bibitem[Myers(2009)]{2009ApJ...700.1609M} Myers, P.~C.\ 2009, \apj, 700, 1609. 
\bibitem[Pineda et al.(2023)]{2023ASPC..534..233P} Pineda, J.~E., Arzoumanian, D., Andre, P., et al.\ 2023, Protostars and Planets VII, 534, 233. 
\bibitem[Shu et al.(1987)]{1987ARA&A..25...23S} Shu, F.~H., Adams, F.~C., \& Lizano, S.\ 1987, \araa, 25, 23.
\bibitem[Sofue(2017)]{2017MNRAS.469.1647S} Sofue, Y.\ 2017, \mnras, 469, 1647. 
\bibitem[Sofue(2021)]{2021PASJ...73L..19S} Sofue, Y.\ 2021, \pasj, 73, L19. 
\bibitem[Sofue(2023)]{2023MNRAS.525.4540S} Sofue, Y.\ 2023, \mnras, 525, 4540. 
\bibitem[Sofue(2023)]{2023Ap&SS.368...74S} Sofue, Y.\ 2023, \apss, 368, 74. 
\bibitem[Sofue \& Nakanishi(2016)]{2016PASJ...68...63S} Sofue, Y. \& Nakanishi, H.\ 2016, \pasj, 68, 63. 
\bibitem[Sofue \& Nakanishi(2017)]{2017PASJ...69...19S} Sofue, Y. \& Nakanishi, H.\ 2017, \pasj, 69, 19.
\bibitem[Solomon et al.(1987)]{1987ApJ...319..730S} Solomon, P.~M., Rivolo, A.~R., Barrett, J., et al.\ 1987, \apj, 319, 730. 
\bibitem[Spilker et al.(2021)]{2021A&A...653A..63S} Spilker, A., Kainulainen, J., \& Orkisz, J.\ 2021, \aap, 653, A63. 
\bibitem[Tokuyama et al.(2019)]{2019PASJ...71S..19T} Tokuyama, S., Oka, T., Takekawa, S., et al.\ 2019, \pasj, 71, S19. 
\bibitem[Umemoto et al.(2017)]{2017PASJ...69...78U} Umemoto, T., Minamidani, T., Kuno, N., et al.\ 2017, \pasj, 69, 78. 
\bibitem[Wong \& Blitz(2002)]{2002ApJ...569..157W} Wong, T. \& Blitz, L.\ 2002, \apj, 569, 157. 

\end{thebibliography}
\end{document}